\documentclass[
reprint,
groupedaddress,
amsmath,
amssymb,
longbibliography,twocolumn
]{revtex4-2}

\usepackage{amsmath}
\usepackage{mathtools}
\usepackage{amssymb}
\usepackage{graphicx}
\usepackage{bm}
\usepackage[colorlinks, linkcolor=blue, citecolor=blue, urlcolor=blue,breaklinks=true]{hyperref}
\usepackage{color,dsfont,multirow,booktabs}
\usepackage{braket}
\usepackage{tabularx}
\usepackage{lipsum,booktabs}
\usepackage{float}

\usepackage{ulem}

\begin{document}

\date{\today}

\title{Path-entangled radiation from kinetic inductance amplifier}

\author{Abdul Mohamed and Shabir Barzanjeh}
\email{shabir.barzanjeh@ucalgary.ca}
\address{Department of Physics and Astronomy, University of Calgary, Calgary, AB T2N 1N4 Canada}

\begin{abstract}
Continuous variable entangled radiation, known as Einstein-Podolsky-Rosen (EPR) states, are spatially separated quantum states with applications ranging from quantum teleportation and communication to quantum sensing. The ability to efficiently generate and harness EPR states is vital for advancements of quantum technologies, particularly in the microwave domain. Here, we introduce a kinetic inductance quantum-limited amplifier that generates stationary path-entangled microwave radiation. Unlike traditional Josephson junction circuits, our design offers simplified fabrication and operational advantages. By generating single-mode squeezed states and distributing them to different ports of a microwave resonator, we deterministically create distributed entangled states at the output of the resonator. In addition to the experimental verification of entanglement, we present a simple theoretical model using a beam-splitter picture to describe the generation of path-entangled states in kinetic inductance superconducting circuits. This work highlights the potential of kinetic inductance parametric amplifiers, as a promising technology, for practical applications such as quantum teleportation, distributed quantum computing, and enhanced quantum sensing. Moreover, it can contribute to foundational tests of quantum mechanics and advances in next-generation quantum information technologies.
\end{abstract}

\maketitle

Quantum entanglement, a fundamental aspect of quantum mechanics, has numerous applications in quantum information processing \cite{RevModPhys.81.865,gisin2014quantum, Ladd2010}, communication \cite{couteau2023applications, Hanson}, precision measurements, and sensing  \cite{Aslam2023, Pirandola2018}. Therefore, the efficient generation of entangled states, such as spatially separated entangled photon pairs known as Einstein-Podolsky-Rosen (EPR) states or two-mode squeezed states, is key for further development of quantum technology and computing \cite{Lvovsky2009}. These states have been used in several experiments to test the foundations of quantum mechanics, such as demonstrating violations of Bell inequalities \cite{PhysRevLett.23.880, PhysRevLett.85.4418, Hensen2015Loophole-free, Storz2023}, and in practical applications like quantum cryptography and quantum teleportation \cite{Guerreiro2016Demonstration,Tashima2008Local}. The robustness of EPR states against loss makes them a valuable asset for long-distance quantum communication and quantum networks~\cite{Andersen2016}. These states also offer unique possibilities in remote quantum sensing and metrology \cite{Lloyd, Barzanjeh, PhysRevLett.114.080503, 2310.07198, Assouly2023}, imaging \cite{Taylor2013, Ono2013An}, and tomography \cite{PhysRevResearch.5.023170}. 

In the optical domain, EPR states are typically generated using nondegenerate optical parametric amplifiers \cite{PhysRevD.26.1817, Lvovsky2009, RevModPhys.82.1155}. In this process, a pump photon passing through a parametric amplifier down-converted into idler and signal photons with lower energy, creating spatially separated two-mode squeezed states~\cite{Heidmann1987, Ou1992, villar2005, Dutt2015} or photon pairs \cite{Kumar:13, ma2017silicon, 2308.11451}. In the microwave frequencies, Josephson parametric amplifiers, operating as nonlinear media \cite{PhysRevApplied.13.024015, PhysRevApplied.13.024014, PhysRevLett.113.110502, PhysRevA.39.2519, Renger2021, JTWPA, JPA_Review, Esposito}, can generate both single-mode~\cite{Yurke1988, Castellanos2008, Mallet2011, Malnou2017} and two-mode~\cite{Eichler2011a, Flurin2012, Menzel2012, Ku2015, Flurin2015, Chan2017} squeezed states. In addition to the relatively complex fabrication process, a major challenge in utilizing Josephson junction circuits for generating microwave entanglement is their sensitivity to stray magnetic fields and the necessity for low-temperature operation. Other methods, such as using electromechanical resonators ~\cite{Barzanjeh2019}, can also produce entanglement between propagating microwave radiations, but they face similar challenges.

An alternative approach to overcoming these limitations is to develop quantum-limited amplifiers (as a source of entanglement), using kinetic inductance superconducting thin films, such as Niobium Titanium Nitride (NbTiN). The inherent nonlinearity of kinetic inductance materials allows much simpler fabrication and chip design compared to Josephson junction amplifiers. Recently, significant progress has been made in developing kinetic inductance amplifiers that achieve very large gain and noise levels approaching the quantum limit \cite{Stannigel, HoEom2012, 10.1063/1.4980102, PhysRevApplied.13.024056, PRXQuantum.2.010302, PhysRevApplied_jarrydPla, PhysRevApplied.21.024011, 2311.11496}. Unlike Josephson junction circuits, the fabrication of these amplifiers is straightforward. Additionally, they demonstrate greater resilience to strong magnetic fields \cite{PhysRevApplied.5.044004, 10.1063/1.4931943, PRXQuantum.4.010322,PhysRevApplied.19.034024} and can operate at higher temperatures exceeding $5$ K \cite{2311.11496}.
While much of the research in this area has focused on amplification or very recently on single-mode vacuum squeezing \cite{PhysRevApplied.19.034024, Vaartjes2024} for weak signal amplification, utilizing kinetic inductance quantum limited amplifiers to deterministically generate spatially separated entangled states has not been achieved so far. 

 \begin{figure*}[t]
    \centering 
     \includegraphics[width=1\linewidth]{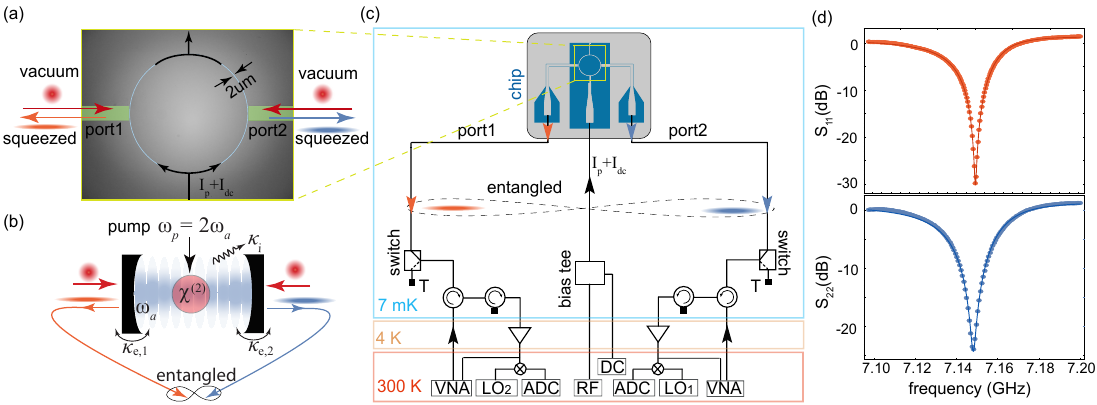
    }
      \caption{ (a) A ring resonator with a diameter of $1.2$ mm and a width of $2$ $\mu m$ is capacitively coupled to two transmission lines on the left and right sides. A separate line running through the top and bottom supplies the dc-current ($I_\text{dc}$) and pump current ($I_\text{p}$) via a bias tee to the system. By properly driving the system, we can generate a squeezed state inside the resonator. (b) A two-sided Fabry-Perot cavity has a resonance frequency $\omega_a$, extrinsic damping rates $\kappa_{\text{e,1}}$ and $\kappa_{\text{e,2}}$, and a dissipation rate $\kappa_\text{i}$. The cavity contains a $\chi^{(2)}$ nonlinear medium. By pumping the cavity at $\omega_p = 2\omega_a$, vacuum squeezing can be generated inside the cavity, resulting in correlations and entanglement between the propagating signals at the outputs of the resonator.  (c) Circuit diagram of the measurement chain on both sides of the resonator. A Vector Network Analyzer (VNA) is used for measuring the resonator's reflection properties. Squeezing is generated by pumping the system at $\omega_p = 2\omega_a$. Microwave signals emitted from the system are amplified using High Electron Mobility Transistor amplifiers (HEMT) at the 4K stage of the dilution refrigerator and room temperature amplifiers, then down-converted using IQ mixers and a local oscillator (LO). The down-converted signals are subsequently filtered and digitized using a two-channel analog-to-digital converter (ADC). The device connects via two 50 cm long copper cables to radial switches at the 7mK stage, facilitating switching between the sample and 50$\Omega$ loads for noise calibration purposes. (c) The reflection parameters $S_{11}$ and $S_{22}$ of the resonator measured at a fixed current $I_{\text {dc}}=1.575$ mA corresponding to a frequency $\omega_0/2\pi=7.14$ GHz.} 
       \label{Fig1}
\end{figure*}
In this paper, we utilize a NbTiN-based Kinetic Inductance Parametric Amplifier (KIPA) to generate stationary emission of path-entangled microwave radiation. The entanglement is generated by the kinetic inductance characteristic of the superconducting film, with the total inductance described as $L_k(I)=L_0[1+(\frac{I}{I^*})^2]$ \cite{Pippard1, Pippard2, Annunziata_2010, Zmuidzinas}, where $L_0$ represents the kinetic inductance of the film in the absence of current. The characteristic current, $I^*$, is proportional to the film's critical current and serves as a quantitative measure of the film's responsiveness to the applied current, $I$. As shown in Fig. \ref{Fig1}a, our design incorporates a nonlinear ring resonator with a resonance frequency $\omega_a$, which is capacitively coupled to two transmission lines for probing and reading out the circuit, thereby forming a two-sided nonlinear resonator. This system is analogous to a two-sided Fabry-Perot cavity containing a $\chi^{(2)}$ nonlinear medium, as illustrated schematically in Fig. \ref{Fig1}b. Our two-port resonator design provides a simple method for generating squeezing. However, as we will explain below, this design restricts the squeezing to no more than 3 dB at each port of the system.

By driving the system at $\omega_p = 2\omega_a$, a squeezed state is created inside the cavity and distributed to both ports, generating cross-correlation and thus entanglement between the outputs of the resonators. In the KIPA design, the $\chi^{(2)}$ nonlinearity arises from the kinetic inductance property of the superconducting films. The nonlinearity's strength can be controlled by applying a dc-current ($I_\text{dc}$) to the circuit, resulting in
\begin{equation}
    L_k(I)=L_0\Big[1+\Big(\frac{I_\text{dc}}{I_*}\Big)^2+\frac{2I_\text{rf}I_\text{dc}}{I_*^2}+\Big(\frac{I_\text{rf}}{I_*}\Big)^2\Big],
    \label{Kin}
\end{equation}
where $I=I_\text{dc}+I_\text{rf}$ and $I_\text{rf}$ is the current of the microwave signal. Combining the dc-current and a strong microwave pump with current amplitude $I_\text{p}$ leads to the squeezing Hamiltonian \cite{2311.11496}
\begin{equation} \label{Ham1}
    H=\Delta a^\dagger a+\frac{g}{2} (a^2e^{-i\phi_{p}}+a^{\dagger2}e^{i\phi_{p}}),
\end{equation}
where $a$ is the annihilation operator of the ring resonator, $g=g_0|\alpha_p|$ is the 3WM parameter, in which $|\alpha_p|$ represents the amplitude of the pump with phase $\phi_p$. The Hamiltonian above has been written in a reference frame rotating at $\omega_p/2=\omega_a-\Delta$.

The emergence of squeezing and entanglement can be directly seen by solving the quantum Langevin equations of motion along with using input-output theory for a two-sided cavity. This gives the output operators $a_\mathrm{out,1(2)}$ in terms of the input noise operators $a_\mathrm{e,1(2)}$ at port 1 (2) of the resonator 
\begin{eqnarray}\label{aout0}
a_\mathrm{out,1(2)}&=&\Big[\mathcal{G}_{S,1(2)}a_\text{e,1(2)}+\mathcal{G}_{I,1(2)}a_\text{e,1(2)}^\dagger\Big]\nonumber\\
&+&\sqrt{\frac{\kappa_\text{e,2(1)}}{\kappa_\text{e,1(2)}}}\Big[(\mathcal{G}_{S,1(2)}+1)a_\text{e,2(1)}+\mathcal{G}_{I,1(2)}a_\text{e,2(1)}^\dagger \Big]\nonumber\\
&+&\sqrt{\frac{\kappa_\text{i}}{\kappa_\text{e,1(2)}}}\Big[(\mathcal{G}_{S,1(2)}+1)a_\text{i}+\mathcal{G}_{I,1(2)}a_\text{i}^\dagger \Big],
\end{eqnarray}
 where $\kappa_\text{e,1(2)}$ is the extrinsic coupling rate between the resonator and port $1(2)$ while \(\kappa_\mathrm{i}\) shows the intrinsic damping rate of the ring resonator with corresponding noise operator $a_\text{i}$. In Eq. (\ref{aout0}), the first bracket on the right-hand side of $a_\mathrm{out,1}$ describes the single-mode squeezing of the resonator seen at port 1. The second and third brackets represent the contributions of port 2 and internal loss, respectively. These terms act as noise terms with rates $\frac{\kappa_\text{e,2}}{\kappa_\text{e,1}}$ and $\frac{\kappa_\text{i}}{\kappa_\text{e,1}}$, which reduce the level of measurable squeezing at port 1. The same interpretation also applies when measuring the output of port 2, where port 1 serves as a loss channel. For $\Delta=\omega=0$, the system's gain parameters are given by $\mathcal{G}_{S,1(2)}=\Big(\frac{2\eta_{1(2)}}{1-\mathcal{C}}\Big)-1, \,\,
   \mathcal{G}_{I,1(2)}=\Big(\frac{2\eta_{1(2)} \sqrt{\mathcal{C}} }{1-\mathcal{C}}\Big)e^{i(\phi_p-\frac{\pi}{2})}$, 
where $\mathcal{C}=4g^2/\kappa^2$ is the cooperativity of the parametric amplification and $\eta_{1(2)}=\frac{\kappa_\text{e,1(2)}}{\kappa}$ is the ratio of the extrinsic damping rate to the total coupling rate $\kappa=\kappa_\text{e,1}+\kappa_\text{e,2}+\kappa_\text{i}$.
 Note that for a single port lossless resonator $\eta_1\approx 1$ and $\eta_2\approx 0$, thus $
a_\mathrm{out}=\mathcal{G}_{S}a_\text{e}+\mathcal{G}_{I}a_\text{e}^\dagger$
 with $|\mathcal{G}_{S}|^2-|\mathcal{G}_{I}|^2=1$, as expected for an ideal single-mode parametric amplifier \cite{PhysRevD.26.1817, RevModPhys.82.1155}.

Eq. (\ref{aout0}) possesses the typical structure of a squeezing operator, resulting in quadrature squeezing below the vacuum noise. To demonstrate this, we introduce the quadrature operators $P_\mathrm{out}=(a_\mathrm{out}-a_\mathrm{out}^\dagger)/\sqrt{2}i$ and $X_\mathrm{out}=(a_\mathrm{out}+a_\mathrm{out}^\dagger)/\sqrt{2}$ along with their variance by setting $\phi_p=-\pi/2$, 
\begin{equation}\label{squeezing}
    \Delta X_\mathrm{out,1(2)}^2=\frac{1}{2}\Big(1-\frac{2\eta_{1(2)}}{1+\sqrt{\mathcal{C}}}\Big)^2+n_{\text{noise,1(2)}},
\end{equation}
where $n_{\text{noise,1(2)}}=\frac{2\eta_{1(2)}[1-\eta_{1(2)}]}{(1+\sqrt{\mathcal{C}})^2}$ is the added noise due to internal losses of the system and the presence of port 2 (1). The first term of Eq. (\ref{squeezing}) shows single-mode squeezing for $\mathcal{C}>0$. This becomes more evident when considering an ideal single-port resonator where $\eta_1= 1$ and $\eta_2=0$, resulting in single-mode squeezing of the quadrature $X_\mathrm{out}$ with a variance $\Delta X_\mathrm{out}^2=\frac{1}{2}\Big(\frac{1-\sqrt{\mathcal{C}}}{1+\sqrt{\mathcal{C}}}\Big)^2=\frac{e^{-2r}}{2}$, see Supplementary Information. Here the squeezing parameter $r$ is defined as $r=2\text{tanh}^{-1}(\sqrt{\mathcal{C}})$.

The squeezing distribution in the current two-port resonator system can effectively be compared to a lossy beam splitter, with transmission and reflection parameters proportional to $\eta_1$ and $\eta_2$. This analogy becomes more clear for a two-port lossless ($\kappa_\mathrm{i}=0$) symmetric resonator ($\eta_1=\eta_2=\frac{1}{2}$) in which Eq. (\ref{squeezing}) reduces to

\begin{equation}\label{squeezing2}
    \Delta X_\mathrm{out,1(2)}^2=\frac{1+\mathcal{C}}{2(1+\sqrt{\mathcal{C}})^2},
\end{equation}
At $\mathcal{C}=1$, the output of both ports reaches its maximum squeezing, $\Delta X_\mathrm{out,1}^2=\Delta X_\text{out,2}^2=\frac{1}{4}$, corresponding to $3$ dB squeezing below vacuum. The symmetric lossless resonator effectively acts as an ideal balanced beam splitter, with a squeezed state in one input port and a vacuum state in the other port. In this scenario, the maximum achievable squeezing at the output ports of the beam splitter is $3$ dB. Note that for $\mathcal{C}\gg1$,  $\Delta X_\mathrm{out,1(2)}^2=\frac{1}{2}$. 

The generation and distribution of squeezing in the two-sided resonator can also lead to entanglement between the output fields of the system mainly due to the generation of quadratures cross-correlation $\langle X_\text{out,1}X_\text{out,2}\rangle= -\frac{2\sqrt{\mathcal{C}\eta_{1}\eta_{2}}}{({1+\sqrt{\mathcal{C}}})^2}$.
This behavior aligns with the predictions of the beam splitter model: injecting a single-mode squeezed state into one port and a vacuum state into the other port results in two correlated squeezed states at the output ports of the beam splitter. The outputs of the resonator are entangled if the parameter \cite{PhysRevLett.84.2722}
\begin{equation}
    \Delta_\text{EPR}:=\Delta X_{\text{+}}^{2}+\Delta P_{\text{--}}^{2}<1
\end{equation}
where  $X_{\text{+}}=\frac{1}{\sqrt{2}}(X_\text{out,1}+X_\text{out,2}),
P_{\text{--}}=\frac{1}{\sqrt{2}}(P_\text{out,1}-P_\text{out,2})$ are the EPR quadratures. For a lossless symmetric two-sided resonator, $\eta_1=\eta_2=\frac{1}{2}$, we obtain $\Delta_\text{EPR}=\frac{1+\mathcal{C}}{(1+\sqrt{\mathcal{C}})^2}$. This expression indicates that the output of the system is always entangled, $\Delta_\text{EPR}<1$, for $\mathcal{C}>0$, and approaches the vacuum state, $\Delta_\text{EPR}=1$, for $\mathcal{C}\gg1$. The maximum entanglement, however, is achieved at $\mathcal{C}=1$, corresponding to $\Delta_\text{EPR}=\frac{1}{2}$ or again $3$ dB squeezing.\newline

We experimentally demonstrate the generation of path-entangled emission for split squeezed states at the outputs of a microwave resonator using a kinetic inductance superconducting thin film. The optical image of the circuit and the measurement setup are shown in Fig.~\ref{Fig1}a and b. The circuit includes a ring-resonator with resonance frequency $\omega_a/2\pi=7.147$~GHz that is capacitively coupled to the input transmission lines from the left and right sides. The reflection parameters for both ports $S_{11}$ and $S_{22}$ are shown in Fig~\ref{Fig1}d. Our design is very simple and doesn't require a Bragg mirror or impedance-matching step coupler \cite{Stannigel, HoEom2012, 10.1063/1.4980102, PhysRevApplied.13.024056, PRXQuantum.2.010302}. The ring resonator can be considered as a two-sided resonator with the extrinsic damping rates $\kappa_\text{e,1}/2\pi=19.4$ MHz and $\kappa_\text{e,2}/2\pi=13.2 $ MHz for channels 1 and 2, respectively. The intrinsic damping rate of the resonator is $\kappa_\text{i}/2\pi\approx 3$ MHz. Note that we determine the sheet inductance, \(L_{\square}\), by measuring the resonance frequency of the ring resonator at zero bias current. By performing simulations in Sonnet and adjusting the sheet inductance to match the experimental data, we deduce the sheet inductance of the film to be \(L_{\square} = 30\) pH. This calculation assumes that the film is uniformly evaporated across the entire chip. Consequently, the total inductance corresponding to this sheet inductance is \(L_0 = 251\) nH.

Figure~\ref{Fig1}c shows the experimental setup used to measure squeezing and entanglement from the chip. The output signals from the left and right sides of the ring resonator are amplified using two identical cryogenic high electron mobility transistor (HEMT) amplifiers. Subsequently, the signals are further amplified at room temperature, filtered, down-converted to an intermediate frequency $\omega_\text{if}/2\pi=20$ MHz, and digitized at a sampling rate of 100 MHz by an 8-bit Alazar-tech analog-to-digital converters~(ADC).

The first step to measure non-classical radiation is to evaluate the mean $\langle X\rangle$ and variance $(\Delta X)^2=\langle X^2\rangle-\langle X\rangle^2$ of the quadrature operator $X$ from experimental data. Utilizing the Fast Fourier transform (FFT) and subsequent post-processing of the measured data, we find the quadrature voltages $I$ and $Q$ at channels 1 and 2 of the ADC. These voltages can be used to compute the scaled (unitless) quadratures $X=\frac{I}{\sqrt{2\hbar \omega_a B R G_\text{T}}}$ and $P=\frac{Q}{\sqrt{2\hbar \omega_a B R G_\text{T}}}$ of the generated signal at the outputs of the ring-resonator. Here, $R=50\,\Omega$, and $B=200$ kHz represent the measurement bandwidth set by a digital filter \cite{Menzel2010, Eichler2011a, Barzanjeh2019, Barzanjeh}. The measured quadratures $X$ and $P$ have the same statistical characteristics as the quadrature operators $ X=\frac{\hat a+\hat a^\dagger}{\sqrt{2}}$ and $ P=\frac{\hat a-\hat a^\dagger}{\sqrt{2}i}$. We calibrate the noise properties of both measurement chains and accurately determine the gains $G_\text{T,1}=99.22\,(0.02)$ dB and $G_\text{T,2}=94.02\,(0.02)$ dB with the number of noise quanta $n_{\mathrm{add,1}}=7.51\,(0.03)$, $n_{\mathrm{add,2}}=14.80\,(0.09)$ added by the output chains, see the Supplemental Information. In the gain and added noise calculations, we do not consider the loss of the copper cables, $\mathcal{A}\approx 0.8$ dB, connecting the ring-resonator sample to the cold switch. As a result, the reported squeezing values reflect the signal propagation over a $0.5$-meter lossy cable. After the cold switch, the initially created squeezed state is affected by amplifier noise, additional losses, and exposure to a thermal bath at elevated temperatures.

\begin{figure}[t]
    \centering 
     \includegraphics[width=1\linewidth]{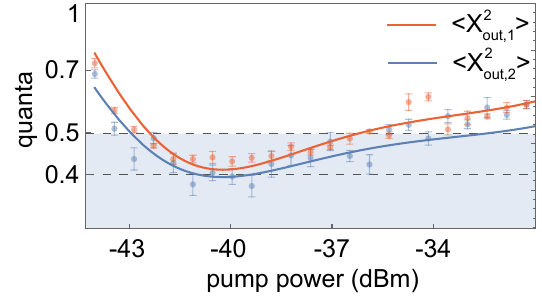}
       \caption{Measured variance of quadratures $X_\text{out,1}(\phi)$ and $X_\text{out,2}(\phi)$ (dots) for ports 1 and 2 of the resonator at an optimal detector angle $\phi$ for different pump powers, fitted to the theoretical model (solid lines). The blue area indicates the region of vacuum squeezing. Error bars represent the standard deviation of the mean from 5 independent measurements of both quadratures, each comprising 115,200 values.}
       \label{Fig2}
\end{figure}

The squeezing of the propagating output states can be fully characterized from the variances of the rotated quadrature $X(\phi)$  when pumps are on and off, i.e., $\langle X^2 
(\phi)\rangle=\langle X^2\rangle_{\mathrm{on}}-\langle X^2\rangle_{\mathrm{off}}+\xi$, where the parameter $\phi$ shows a rotation of the detector phase, implemented during post-processing. Here, $\xi=\frac{1}{2} \mathrm{coth}\,\hbar \omega_a/(k_B T)\approx 0.5$ shows the input quantum noise at temperature $T$, and the brackets $\langle..\rangle$ indicate an average over the $576000$ measurements when pump were on or off. 
Figure~\ref{Fig2} shows the measured quadrature squeezing for outputs 1 and 2, $\langle X_{1(2)}^2\rangle$, at the optimal angle $\phi$ as a function of drive power $P_p$. The parameter $\langle X^2 \rangle$ remains consistently below 0.5, representing the squeezed nature of the ring resonator's output, reaching 0.7(0.2) dB and 1.0(0.3) dB squeezing for output 1 and 2, respectively. However, at higher pump powers, the squeezing gradually diminishes and eventually disappears. This phenomenon is attributed to finite losses in the superconducting film and the resonator, leading to pump power–dependent heating and resulting in larger variances of the output field. The reported errors represent the statistical errors of the measured means, which surpass the statistical errors and the long-term variation observed in the subtracted noise measurements (with the pump turned off), in addition to the error from the calibration measurements.

\begin{figure}[t]
    \centering 
     \includegraphics[width=1\linewidth]{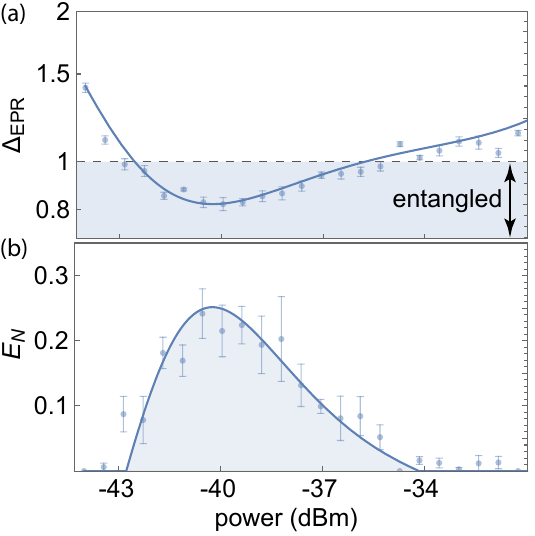}
       \caption{ (a) Measured EPR parameters $\Delta_\text{EPR}:=\Delta X_{\text{+}}^{2}+\Delta P_{\text{--}}^{2}$ at different pump powers for optimized detector angles. (b) Logarithmic-negativity $E_N$ as a function of pump power. Solid lines represent the theoretical model fitting the experimental results (dots). Mean values and error bars are obtained in the same manner as in Fig. (\ref{Fig2}).}
       \label{Fig3}
\end{figure}

The appearance of squeezing at each port can indicate the presence of entanglement between the spatially separated ports of the resonator. We confirm this by computing the variance of the EPR quadratures $\langle X_{\text{+}}(\phi)^2\rangle=\frac{1}{2}\langle (X_\text{out,1}(\phi)+X_\text{out,2})^{2}\rangle$ and $\langle P_{\text{--}}(\phi)^2\rangle=\frac{1}{2}\langle (P_\text{out,1}(\phi)-P_\text{out,2})^{2}\rangle$, and calculate the EPR parameter $\Delta_\text{EPR}:=\Delta X_{\text{+}}^{2}+\Delta P_{\text{--}}^{2}$. In Fig. \ref{Fig3}a, we measure parameter $\Delta_\text{EPR}$ at an optimized angle $\phi$ for different pump powers. This figure shows that the outputs of the ring resonator are entangled reaching 0.9(0.2) dB squeezing.

We additionally quantify the amount of entanglement generated by our microwave circuits using standard
 measures in quantum information theory. Specifically, we are interested in the logarithmic-negativity \cite{PhysRevA.65.032314, PhysRevLett.95.090503}, which serves as an upper limit on the amount of distillable entanglement bits produced by the source. The log-negativity $E_N$ is given by
\begin{equation}
    E_{N}=\text{max}[0,-2\text{log}(\zeta^{-})]
\end{equation}
where $\zeta^{-}$ is the smallest partially-transposed symplectic eigenvalue of the covariance matrix defined $ V_{ij}= \frac{\langle X_{i}X_{j}+X_{j}X_{i} \rangle}{2},$
with $X_{ij}=\{X_\text{out,1},P_\text{out,1},X_\text{out,2},P_\text{out,2}\}$, see Supplemental Information. Figure \ref{Fig3}b shows logarithmic negativity versus the pump power with the maximum value $E_N= 0.24 (0.05)$. The log-negativity exhibits similar behavior to the EPR parameter; increasing the pump power causes the entanglement to diminish.

In summary, we have demonstrated the generation of stationary microwave path-entangled radiation using kinetic inductance superconducting quantum circuits for the first time. Compared to Josephson junction amplifiers, our device utilizes a single-layer NbTiN thin film, offering a high fabrication yield. By generating single-mode squeezing and splitting it with a two-port microwave resonator, we demonstrated entanglement generation and distribution between two separate paths. Although this work did not explore resilience to magnetic fields or operation at higher temperatures, high-kinetic inductance NbTiN resonators exhibit exceptional magnetic field compatibility \cite{PhysRevApplied.5.044004, PhysRevApplied.19.034024, PhysRevApplied.21.024011, 2311.07968}. This property makes the devices suitable for integration with spin systems and defect centers \cite{PhysRevX.7.041011}, such as NV centers, enabling precise readout of quantum nodes or entanglement distribution between remote spins in different substrates.

Several improvements can further enhance the efficiency of this system. One approach is to reduce loss and noise by refining the fabrication process or using resonators with better impedance matching. The current design limits squeezing to 3 dB per port. To achieve higher levels of squeezing and consequently larger entanglement, designing two-mode resonance circuits to generate two-mode squeezed states would be beneficial. This device could potentially be used to generate microwave nonclassical emissions at higher temperatures, paving the way for designing sensors with real-world applications in quantum illumination and microwave quantum radars \cite{2310.07198} or accelerometry without the need for dilution refrigerators or operation at milliKelvin temperatures. 

\textbf{Acknowledgments} We thank Jarryd Pla for
 helpful comments and discussions. S.B. acknowledges funding by the Natural Sciences and Engineering Research Council of Canada (NSERC) through its Discovery Grant and Quantum Alliance Grant, funding and advisory support provided by Alberta Innovates (AI) through the Accelerating Innovations into CarE (AICE) -- Concepts Program, support from Alberta Innovates and NSERC through Advance Grant project, and Alliance Quantum Consortium. This project is funded [in part] by the Government of Canada. Ce projet est financé [en partie] par le gouvernement du Canada.

\bibliography{BarzanjehGroup}

\newpage
\onecolumngrid
\appendix
\begin{center}
\textbf{\large Supplementary Materials}
\end{center}

\subsection{Hamiltonian and theoretical model }

The Hamiltonian describing a two-port Fabry–Pérot cavity containing nonlinearity  $\chi^{(2)}$  and resonance frequency $\omega_a$ is given by,
\begin{equation} \label{Hamsinglemode000}
    H=\omega_a a^\dagger a+\frac{g_0}{2} (a^2 a_p^\dagger+a^{\dagger2}a_p),
\end{equation}
This Hamiltonian represents a Three-Wave-Mixing (3WM) process, where a pump photon at frequency \(\omega_p\) is annihilated, resulting in the creation of two photons at the resonance frequency \(\omega_a\). In a strong pump regime, we can treat the pump operator as a classical variable, \(\alpha_p = |\alpha_p| e^{-i\phi_p} e^{-i\omega_p t}\), where \(|\alpha_p|\) is the amplitude of the pump and \(\phi_p\) is its phase. This approximation along with moving to a rotating frame with respect to the frequency of the pump allows us to rewrite the Hamiltonian (\ref{Hamsinglemode000}) 

\begin{equation}
\label{HamiltonianSup}
H=\Delta a^{\dagger}a+\frac{g}{2}(a^{2} e^{i \phi_{p}}+a^{\dagger^2}e^{-i \phi_{p}})
\end{equation}
where $g=g_0|\alpha_p|$ and $\omega_p/2=\omega_a-\Delta$.

By using the quantum Langevin equations of motion, we derive the following set of coupled equations that describe the system's dynamics for a two-sided nonlinear cavity
\begin{eqnarray}
 \dot{a}&=&-\Big(i\Delta+\frac{\kappa}{2}\Big)a-ig e^{i\phi_p}a^\dagger+\sum_{j=1,2}\sqrt{\kappa_{\mathrm{e},j}}a_{\mathrm{e},j}+\sqrt{\kappa_\mathrm{i}}a_{\mathrm{i}},\nonumber\\
  \dot{a}^\dagger&=&\Big(i\Delta-\frac{\kappa}{2}\Big)a^\dagger+ig e^{-i\phi_p}a+\sum_{j=1,2}\sqrt{\kappa_\mathrm{e,j}}a_{{\mathrm{e},j}}a_{\mathrm{e},j}^\dagger+\sqrt{\kappa_\mathrm{i}}a_{\mathrm{i}}^\dagger,
     \label{eqmotionSI}
\end{eqnarray}
where \(\kappa = \kappa_{\mathrm{e},1} + \kappa_{\mathrm{e},2} + \kappa_{\mathrm{i}}\) represents the total damping rate of the resonator. These equations incorporate quantum noise affecting input fluctuations for the resonator ($a_\mathrm{e,1(2)}$ with extrinsic damping rate $\kappa_\mathrm{e,1(2)}$) and the intrinsic losses in the resonator mode ($a_\mathrm{i}$ with intrinsic damping rate $\kappa_\mathrm{i}$). The correlations for these noises are defined as follows:
\begin{eqnarray}
    \langle a_{\mathrm{e(i)}} (t) a_{\mathrm{e(i)}}^\dagger (t')\rangle&=&\langle a_{\mathrm{e(i)}}^\dagger  (t) a_{\mathrm{e(i)}}(t')\rangle+\delta (t-t')\nonumber\\
    &=&(\bar{n}_{\mathrm{e(i)}}+1)\delta(t-t'),
\end{eqnarray}
where $\delta (t)$ is the Dirac delta function and $\bar{n}_{\mathrm{e(i)}}$ are the Planck-law thermal occupancies of microwave mode (bath).
 By moving to the Fourier domain, where \(\dot{O} \rightarrow -i\omega\), and applying input-output theory, we can find the output operators at ports 1 and 2 of the cavity \(a_{\text{out, 1(2)}}\)

\begin{eqnarray}
\label{aout0Sup}
a_\mathrm{out,1(2)}&=&\Big[\mathcal{G}_{S,1(2)}a_\text{e,1(2)}+\mathcal{G}_{I,1(2)}a_\text{e,1(2)}^\dagger\Big]
+\sqrt{\frac{\kappa_\text{e,2(1)}}{\kappa_\text{e,1(2)}}}\Big[(\mathcal{G}_{S,1(2)}+1)a_\text{e,2(1)}+\mathcal{G}_{I,1(2)}a_\text{e,2(1)}^\dagger \Big]\nonumber\\
&+&\sqrt{\frac{\kappa_\text{i}}{\kappa_\text{e,1(2)}}}\Big[(\mathcal{G}_{S,1(2)}+1)a_\text{i}+\mathcal{G}_{I,1(2)}a_\text{i}^\dagger \Big],
\end{eqnarray}
with idler and signal gain parameters
 \begin{eqnarray}
   \mathcal{G}_{S,1(2)}(\omega)&=&\frac{\eta_{1(2)} \kappa \Big[\frac{\kappa}{2}-i(\omega+\Delta)\Big]}{\Delta^2-g^2+(i\omega-\frac{\kappa}{2})^2}-1,\nonumber\\
   \mathcal{G}_{I,1(2)}(\omega)&=&\frac{-i\eta_{1(2)} \kappa g e^{i\phi}}{\Delta^2-g^2+(i\omega-\frac{\kappa}{2})^2},
 \end{eqnarray}
where $\eta_{1(2)}=\frac{\kappa_\text{e,1(2)}}{\kappa}$ describes the waveguide-resonator coupling. The Eq. (\ref{aout0Sup}) fulfills the bosonic commutation relation for the output field operators $[a_\mathrm{out,j} (\omega), a_\mathrm{out,k}^{\dagger}(\omega')]=\delta(\omega-\omega')\delta_\text{jk}$, resulting in
\begin{equation}\label{commu}
    \frac{|\mathcal{G}_{I,1(2)}|^2 }{{\eta_{1(2)}}}=|\mathcal{G}_{S,1(2)}|^2+\Big(\frac{1-\eta_{1(2)}}{\eta_{1(2)}}\Big)|\mathcal{G}_{S,1(2)}+1|^2-1,
\end{equation}
where for a lossless and single port resonator for example $\eta_1=1$ and $\eta_2=0$, we get $|\mathcal{G}_{S,1}|^2-|\mathcal{G}_{I,1}|^2=1$. 

\subsection{Squeezing and cross-correlation }

By setting $\Delta=\omega=0$ and defining $\mathcal{C}=4g^2/\kappa^2$, we can simplify the gain parameters to

\begin{eqnarray}\label{gainS}
   \mathcal{G}_{S,1(2)}&=&\frac{2\eta_{1(2)}}{1-\mathcal{C}}-1,\nonumber\\
   \mathcal{G}_{I,1(2)}&=&-\frac{2i\eta_{1(2)} \sqrt{\mathcal{C}} }{1-\mathcal{C}}e^{i\phi_p},
 \end{eqnarray}

as presented in the main text. 

Using Eq. (\ref{aout0Sup}) we can calculate the variance of quadratures  
\begin{eqnarray}
X_{\mathrm{out,1(2)}}=\frac{a_{\mathrm{out,1(2)})}+a_{\mathrm{out,1(2)}}^{\dagger}}{\sqrt{2}} \\
    P_{\mathrm{out,1(2)}}=\frac{a_{\mathrm{out,1(2)}}-a_{\mathrm{out,1(2)}}^{\dagger}}{\sqrt{2}i}
\end{eqnarray}
For this we assume the input noise operators from the waveguides are all in vacuum $\bar{n}_{\mathrm{e,1(2)}}=0$, thus the variances of quadratures are given by
\begin{eqnarray}
    \Delta X_{\mathrm{out,1(2)}}^{2}(\phi_p)&=& \braket{X_{1(2)}^{2}}-\braket{X_{1(2)}}^2=\frac{1}{2}+
\frac{|\mathcal{G}_{I,1(2)}|^2}{\eta_{1(2)}}+\frac{\kappa_\text{i}\bar{n}_{\mathrm{i}}}{\kappa_{\text{e,1(2)}}} \Big[|\mathcal{G}_{S,1(2)}+1|^2+|\mathcal{G}_{I,1(2)}|^2\Big]\\
&+&\Big(|\mathcal{G}_{I,1(2)}||\mathcal{G}_{S,1(2)}|+\frac{\eta_{2(1)}}{\eta_{1(2)}}|\mathcal{G}_{I,1(2)}||\mathcal{G}_{S,1(2)}+1|+\frac{\kappa_\text{i}(2\bar{n}_{\mathrm{i}}+1)}{\kappa_{\text{e,1(2)}}} |\mathcal{G}_{I,1(2)}||\mathcal{G}_{S,1(2)}+1|\Big)\text{sin}(\phi_p),\nonumber
\end{eqnarray}
and
\begin{eqnarray}
    \Delta P_{\mathrm{out,1(2)}}^{2}(\phi_p)&=& \braket{P_{1(2)}^{2}}-\braket{P_{1(2)}}^2= \frac{1}{2}+
\frac{|\mathcal{G}_{I,1(2)}|^2}{\eta_{1(2)}}+\frac{\kappa_\text{i}\bar{n}_{\mathrm{i}}}{\kappa_{\text{e,1(2)}}} \Big[|\mathcal{G}_{S,1(2)}+1|^2+|\mathcal{G}_{I,1(2)}|^2\Big]\\
&-&\Big(|\mathcal{G}_{I,1(2)}||\mathcal{G}_{S,1(2)}|+\frac{\eta_{2(1)}}{\eta_{1(2)}}|\mathcal{G}_{I,1(2)}||\mathcal{G}_{S,1(2)}+1|+\frac{\kappa_\text{i}(2\bar{n}_{\mathrm{i}}+1)}{\kappa_{\text{e,1(2)}}} |\mathcal{G}_{I,1(2)}||\mathcal{G}_{S,1(2)}+1|\Big)\text{sin}(\phi_p),\nonumber
\end{eqnarray}

For $\Delta=\omega=0$ and $\bar{n}_{\mathrm{i}}=0$, the $X-$ and $P$-quadrature variances (at optimum phases $\phi_p=\pm \pi/2$) reduce to
\begin{equation}
   \Delta X_\mathrm{out,1(2)}^2|_{\phi_p\rightarrow-\pi/2}= \Delta P_\mathrm{out,1(2)}^2|_{\phi_p\rightarrow\pi/2}=\frac{1}{2}\Big(1-\frac{2\eta_{1(2)}}{1+\sqrt{\mathcal{C}}}\Big)^2+\frac{2\eta_{1(2)}(1-\eta_{1(2)})}{(1+\sqrt{\mathcal{C}})^2},
   \label{vari}
\end{equation}

as explained in the main text. 

The generated squeezing creates a cross-correlation between the outputs of the system, resulting in the following quadrature correlations
\begin{eqnarray}
\langle X_\mathrm{out,1} X_\mathrm{out,2}\rangle &=&\frac{1}{2}\langle (a_\mathrm{out,1}a_\mathrm{out,2}+a_\mathrm{out,1}^{\dagger}a_\mathrm{out,2}^{\dagger}+a_\mathrm{out,1}a_\mathrm{out,2}^{\dagger}+a_\mathrm{out,1}^{\dagger}a_\mathrm{out,2})\rangle=-\frac{2\sqrt{\mathcal{C}\eta_{1}\eta_{2}}}{({1+\sqrt{\mathcal{C}}})^2},\nonumber\\
\langle P_\mathrm{out,1} P_\mathrm{out,2}\rangle &=&\frac{1}{2}\langle (a_\mathrm{out,1}a_\mathrm{out,2}+a_\mathrm{out,1}^{\dagger}a_\mathrm{out,2}^{\dagger}-a_\mathrm{out,1}a_\mathrm{out,2}^{\dagger}-a_\mathrm{out,1}^{\dagger}a_\mathrm{out,2})\rangle=\frac{2\sqrt{\mathcal{C}\eta_{1}\eta_{2}}}{({1-\sqrt{\mathcal{C}}})^2}.
\label{pvar}
\end{eqnarray}
for $\phi_p=-\pi/2$, $\Delta=\omega=0$ and $\bar{n}_{\mathrm{i}}=0$. This now allows us to calculate the entanglement parameter 
\begin{eqnarray}
\Delta_\text{EPR}:=\Delta X_{\text{+}}^{2}+\Delta P_{\text{--}}^{2},
\end{eqnarray}

Using the definition of the quadrature $X_{\text{+}}=\frac{1}{\sqrt{2}}(X_\text{out,1}+X_\text{out,2}),
P_{\text{--}}=\frac{1}{\sqrt{2}}(P_\text{out,1}-P_\text{out,2})$, we get
\begin{eqnarray}
\Delta_\text{EPR}= \frac{1}{2}\Big(\sum_{j=1,2}\big[\Delta X_{\mathrm{out},j}+\Delta P_{\mathrm{out},j}\big]+\langle X_\mathrm{out,1} X_\mathrm{out,2}\rangle +\langle X_\mathrm{out,2} X_\mathrm{out,1}\rangle -\langle P_\mathrm{out,1} P_\mathrm{out,2}\rangle -\langle P_\mathrm{out,2} P_\mathrm{out,1}\rangle\Big),
\label{epr}
\end{eqnarray}

By substituting Eqs. (\ref{vari}) and (\ref{pvar}) into Eq. (\ref{epr}), we find
\begin{eqnarray}
\Delta_\text{EPR}= \frac{\mathcal{C}^2+(1-4\sqrt{\eta_1 \eta_2\mathcal{C}})-2\mathcal{C}(1+2\sqrt{\eta_1 \eta_2\mathcal{C}}-2(\eta_1+\eta_2))}{(1-\mathcal{C})^2},
\end{eqnarray}

For a lossless symmetric two-sided resonator, $\eta_1=\eta_2=\frac{1}{2}$, we obtain $\Delta_\text{EPR}=\frac{1+\mathcal{C}}{(1+\sqrt{\mathcal{C}})^2}$.

\subsection{Covariance matrix and logarithmic-negativity }

We can reconstruct the covariance matrix (CM) between the two output channels of the system, which can be expressed as,

\begin{eqnarray}
     V_{ij}= \frac{\langle X_{i}X_{j}+X_{j}X_{i} \rangle}{2},
     \end{eqnarray}
with $X_{ij}=\{X_\text{out,1},P_\text{out,1},X_\text{out,2},P_\text{out,2}\}$. To verify the generation of entanglement in the system we compute 
the logarithmic negativity $E_N=\text{max}[0,-2\text{log}(\zeta^{-})]$ where 
\begin{eqnarray}
\zeta^{-}=\left(\frac{\Delta-\sqrt{\Delta^{2}-4\textbf{det}V}}{2}\right)^{1/2}
\end{eqnarray}
is the smallest partially-transposed symplectic eigenvalue of the covariance in which 
\begin{equation}
    \nonumber
    \Delta=\textbf{det}A+\textbf{det}B-2\textbf{det}C \quad, \quad V=\begin{pmatrix}
        A&C&\\
        C^{T}&B
    \end{pmatrix}
\end{equation}

\subsection{Sample Fabrication}

The fabrication process begins with a high-resistivity intrinsic silicon wafer ($R \geq 20\,k\, \Omega- \text{cm}$) that is 500 micrometers thick, coated with a 10 nm layer of NbTiN, and diced into smaller chips. The chips undergo a cleaning procedure using acetone and IPA, each involving a 5-minute sonication. Following this, the chip's surface is coated with AZ 1529 resist and patterned using optical lithography. The device is then etched using an ICP Reactive Ion Etcher.

\subsection{Packaging and the PCB design}
After the fabrication, the chip is attached to a gold-plated PCB using silver paste. This PCB, made from Rogers AD 1000 laminate with a dielectric thickness of 0.5 mm, is coated with oxygen-free copper on both sides and finished with gold plating. The PCB includes 50-$\Omega$ coplanar waveguides that connect to the external measurement line via a surface-mount mini-SMP microwave connector. At the opposite end, aluminum bond wires connect the PCB to the input and pump/DC ports of the KIPA resonator.

To mitigate unwanted parasitic modes, an array of vias connects the transmission line on the PCB to the top and bottom ground planes. The entire assembly is then housed in an oxygen-free enclosure, forming a 3D cavity with a resonance frequency distant from the sample's operational point. This cavity shields the sample from parasitic noise and enhances its thermalization.

\subsection{Noise calibration}

\begin{figure*}[t]
    \centering 
     \includegraphics[width=0.7\linewidth]{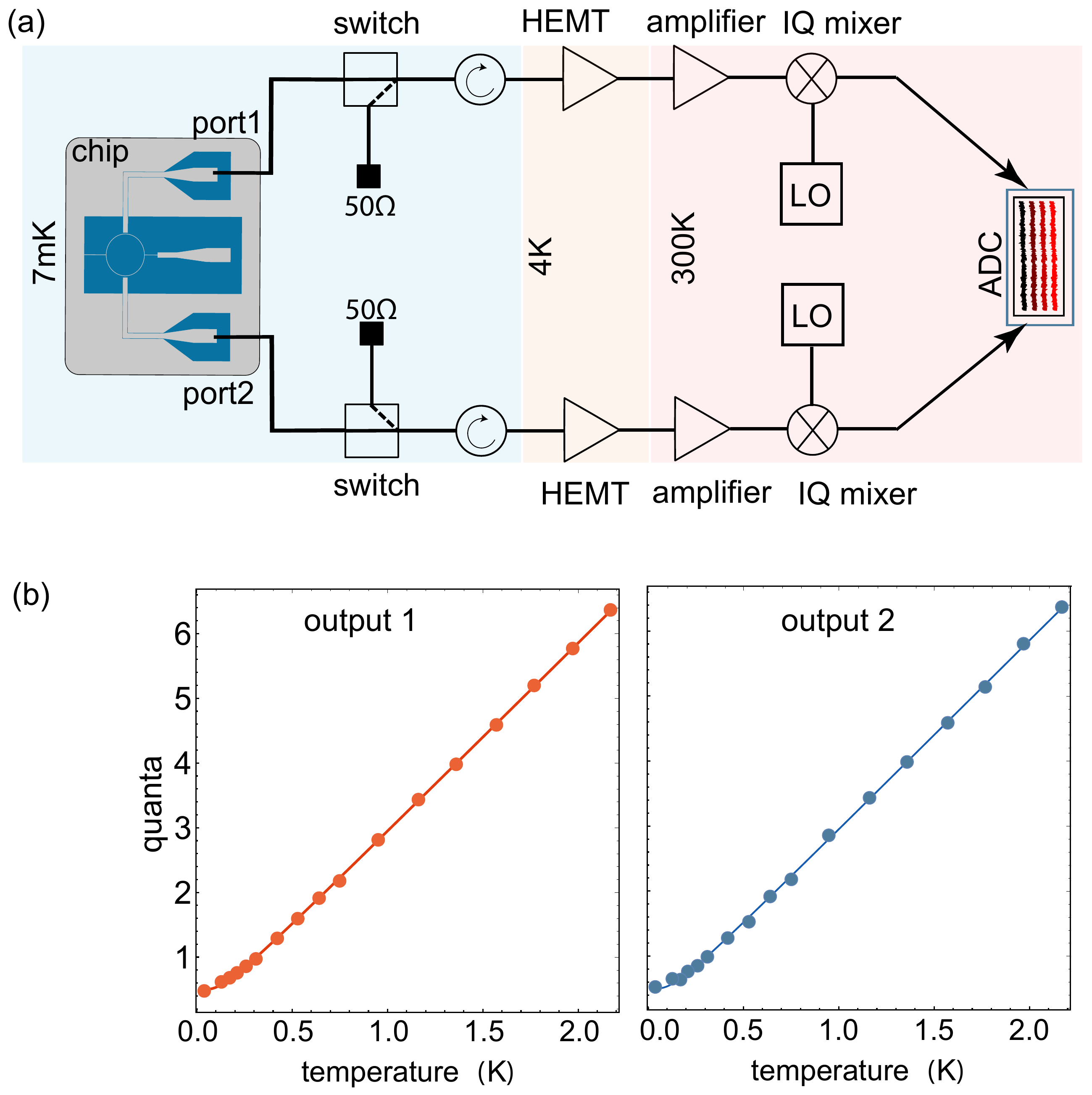}
       \caption{ (a) The schematic of the calibration setup. Two temperature-controlled \(50\,\Omega\) loads are used to generate known broadband thermal noise at the 7 mK stage of the dilution refrigerator. This noise propagates through the same measurement chain used for detecting squeezing and entanglement. At each channel of the measurement chain, the generated noise is amplified using high mobility electron transistors (HEMTs) and room temperature amplifiers, down-converted using IQ mixers, and measured by a two-channel analog-to-digital converter (ADC). (b) By knowing the temperature of the noise calibration tool (the \(50\,\Omega\) load) and fitting the data using the Einstein distribution function, we can accurately determine the gain and added noise for each channel separately. We utilize two cryogenic switches to switch between the sample and the calibration tools. For down-conversion, we use local oscillators (LO) to down converter signals to 20 MHz. } 
       \label{Fig1SI}
\end{figure*}

To accurately verify the entanglement between the two output channels, we first determine the gain and added noise quanta of the measurement chain using two temperature-controlled \(50\,\Omega\) loads \cite{Barzanjeh2019, Barzanjeh, 2312.14900}, see Fig. \ref{Fig1SI}. By adjusting and recording the temperature of the \(50\,\Omega\) load, we can produce predictable thermal noise. This noise is then used to calibrate the measurement chain. The calibration devices are connected to the setup via two 5-cm-long superconducting coaxial cables and a thin copper braid, which provides weak thermal anchoring to the mixing chamber plate, using a Radiall R573423600 latching microwave switch. The noise generated for calibration travels along the measurement chain, is down-converted to 20 MHz, filtered, and then measured with our two-channel ADC. By fitting the noise density in $V^2/\text{Hz}$ to the expected photon number at a given temperature
\begin{equation}
    N_{i}=G_{\text{T,i}}\hbar\omega RB\left(\frac{1}{2}\text{coth}[\hbar\omega/2k_{b}T]\text+n_{\text{add,i}}\right),
\end{equation}
we can accurately determine the gains $G_\text{T,1}=99.22\,(0.02)$ dB and $G_\text{T,2}=94.02\,(0.02)$ dB with the number of noise quanta $n_{\mathrm{add,1}}=7.51\,(0.03)$, $n_{\mathrm{add,2}}=14.80\,(0.09)$ added by the output chains. The confidence intervals are calculated from the standard error of the fits. Here $R=50\,\Omega$ and $B=200$ kHz.

\end{document}